\documentclass[12pt]{article}

\makeatletter
\newcommand*{\indep}{%
\mathbin{%
\mathpalette{\@indep}{}%
}%
}

\newcommand*{\nindep}{%
\mathbin{% % The final symbol is a binary math operator
\mathpalette{\@indep}{\not}% \mathpalette helps for the adaptation
}%
}
\newcommand*{\@indep}[2]{%
\sbox0{$#1\perp\m@th$}% box 0 contains \indep symbol
\sbox2{$#1=$}% box 2 for the height of =
\sbox4{$#1\vcenter{}$}% box 4 for the height of the math axis
\rlap{\copy0}% first \perp
\dimen@=\dimexpr\ht2-\ht4-.2pt\relax
\kern\dimen@
{#2}%
\kern\dimen@
\copy0 % second \perp
}
\makeatother

\usepackage{fancyhdr}
\usepackage{times}
\usepackage{amsmath}
\usepackage{amsthm}
\usepackage{algorithm}
\usepackage{pgf,tikz}
\usepackage{amsfonts}
\usepackage[round]{natbib}
\usepackage{diagbox}
\usepackage{JASA_manu}
\usepackage{graphicx} %插入图片的宏包
{
	\newtheorem{assumption}{Assumption}
	\newtheorem{example}{Example}

	\newtheorem{proposition}{Proposition}

}
\usepackage{bm}
\usepackage{appendix}
\usepackage{tikz}
\usetikzlibrary{arrows,positioning}
\usepackage[colorlinks=true]{hyperref}
\usepackage{algorithm}
\usepackage{amsmath}
\usepackage{graphicx}
\usepackage{fancyhdr}
\usepackage{rotating}
\usepackage{booktabs}
\usepackage{multirow}
\usepackage{chngcntr}
\usepackage{apptools}
\usepackage[english]{babel}
\usepackage[utf8]{inputenc}
\usepackage{algorithm}
\usepackage[noend]{algpseudocode}
\newtheorem{innercustomgeneric}{\customgenericname}
\providecommand{\customgenericname}{}
\newcommand{\newcustomtheorem}[2]{%
  \newenvironment{#1}[1]
  {%
   \renewcommand\customgenericname{#2}%
   \renewcommand\theinnercustomgeneric{##1}%
   \innercustomgeneric
  }
  {\endinnercustomgeneric}
}

\newcustomtheorem{customprop}{Proposition}
\newcustomtheorem{customcond}{Condition}% The preamble here sets up a lot of new/revised commands and
\newcounter{lastnote}

\title{Regression-based Negative Control of  Homophily\\ in Dyadic Peer Effect Analysis}

\author
{Lan Liu$^{1}$  and Eric Tchetgen Tchetgen$^2$ \\
 School of Statistics, University of Minnesota at Twin Cities$^{1}$ \\
Department of Statistics of the Warton School, University of Pennsylvania$^2$}

\date{}

\begin{document}
\maketitle

\begin{abstract}
A prominent threat to causal inference about peer effects over social networks is the presence of homophily bias, that is, social influence between friends and families is entangled with common characteristics or underlying similarities that form close connections. Analysis of social network data has suggested that certain health conditions such as obesity and psychological states including happiness and loneliness can  spread over a network. However, such analyses of peer effects or contagion effects have come under criticism because homophily bias may compromise the causal statement. We develop a regression-based approach which leverages a negative control exposure for identification and estimation of contagion effects on additive or multiplicative scales, in the presence of homophily bias. We apply our methods to evaluate the peer effect of obesity in Framingham Offspring Study.% and find empirical evidence that obesity may spread in mutual relationships but not necessarily in one-sided relationships.
\end{abstract}

%  Please place your key words in alphabetical order, separated
%  by semicolons, with the first letter of the first word capitalized,
%  and a period at the end of the list.
%


%  As usual, the \maketitle command creates the title and author/affiliations
%  display 

%  If you are using the referee option, a new page, numbered page 1, will
%  start after the summary and keywords.  The page numbers thus count the
%  number of pages of your manuscript in the preferred submission style.
%  Remember, ``Normally, regular papers exceeding 25 pages and Reader Reaction 
%  papers exceeding 12 pages in (the preferred style) will be returned to 
%  the authors without review. The page limit includes acknowledgements, 
%  references, and appendices, but not tables and figures. The page count does 
%  not include the title page and abstract. A maximum of six (6) tables or 
%  figures combined is often required.''

%  You may now place the substance of your manuscript here.  Please use
%  the \section, \subsection, etc commands as described in the user guide.
%  Please use \label and \ref commands to cross-reference sections, equations,
%  tables, figures, etc.
%
%  Please DO NOT attempt to reformat the style of equation numbering!
%  For that matter, please do not attempt to redefine anything!
\section{Introduction}

In social network studies, it is of great interest to assess the causal contagion effect of one individual on their social contacts. Classical causal inference was primarily developed within the potential outcome framework and typically involves a ``no interference'' assumption. Recently, causal inference research has extended the classical  potential outcome framework to allow for interference, i.e., that an individual's outcome may be affected by another's exposure \citep{sobel2006randomized,hudgens2008toward,vanderweele2011bounding,tchetgen2012causal,liu2013asymptotic,liu2016inverse}. However, inferring causation from networks remains challenging because correlation in outcomes between individuals with network ties may not only be due to social influence, but also to latent factors that influence network formation. The  phenomenon that individuals tend to associate and bond with persons that they have most in common with  is known as homophily \citep{shalizi2011homophily}.%The  phenomenon of having spurious correlation due to network formation is known as homophily (Vanderweele and An, 2013\nocite{vanderweele2013social}).

Different types of experimental designs and analytic methods have been developed to study network formation or to adjust for homophily bias. For example, Camargo et al. (2010)\nocite{camargo2010interracial} investigated friendship formation among randomly assigned roommates in college and concluded that randomly assigned roommates of different races are as likely to become friends as of the same race. In observational studies, Christakis and Fowler (2007)\nocite{christakis2007spread} explored the spread of obesity to one individual (ego) from their friend or spouse (alter). Specifically, they included in a regression model for ego's BMI, a time-lagged measurement of ego's obesity status, the obesity status of alter, a time-lagged measurement of alter's obesity status and some observed covariates. They found that obesity spreads through social ties. Using the same approach, Christakis and Fowler examined the evidence of social influence for smoking, happiness, loneliness, depression, drug use, and alcohol consumption (Christakis and Fowler 2007, 2008; Fowler and Christakis 2008; Christakis and Fowler 2013\nocite{christakis2007spread}\nocite{christakis2008collective}\nocite{fowler2008estimating}\nocite{christakis2013social}). 

In recent years, published analyses by Christakis and Fowler have come under critical scrutiny. For instance, Shalizi and Thomas (2011)\nocite{shalizi2011homophily} argued that controlling for alter's lagged obesity status may at best only partially account for homophily bias. They pointed out that if the latent factor influencing friendship formation affects current obesity status even after controlling for past obesity status, one may still observe an association between ego's and alter's obesity status using classical regression methods even if alter has no social influence on ego's obesity status. \citet{cohen2009detecting} argued that using the same method as Christakis and Fowler's on traits unlikely to be transmitted over a network such as height, acne and headaches led to the same conclusion that they spread over the network. To account for both unmeasured confounding and homophily, \citet{o2014estimating} leveraged multiple genes in an instrumental variables (IV) approach to identify peer effects under a linear model for the outcome and exposure. They assume that the causal relationship is non-directional and  found a positive causal peer effect of BMI between ego and alter using this IV approach. However, the IV approach requires the exclusion restriction  that none of the genes used to define the IV has a causal effect on any of the unmeasured factors that give rise to formation of social ties, an assumption which may be difficult to justify in social network problems \citep{fowler2009model}.

In this paper, we are also interested in evaluating the person-to-person spread of traits in a social network. We develop an alternative regression-based approach that explicitly accounts for the presence of homophily bias without requiring a valid IV or relying on linear exposure and outcome regression models. Instead of an IV based design, we consider a negative control design that one  observes a variable associated with the unmeasured factor inducing homophily bias, unconfounded with the outcome, and that does not have a direct causal effect on the outcome in view. Such a  variable is formally called a negative control exposure variable. 

Negative control variables have primarily been used in epidemiological applications to detect and sometimes correct for unmeasured confounding \citep{lipsitch2010negative,tchetgen2013control,sofer2016negative,miao2018identifying}.  \cite{elwert2008wives} recently used a negative control exposure to detect homophily bias in the analysis of dyadic data, i.e., data with pairs of two individuals. %and assumptions needed for identification are conflated with assumptions needed for estimation and inference CHECK THAT THIS IS TRUE. 
Specifically, they used the death of an ex-wife as a negative control variable to investigate the ``widowhood effect'', i.e., the effect of the death of a spouse on the mortality of a widow. However, they do not provide a formal counterfactual  approach for inference leveraging a negative control outcome to completely account for homophily bias. Partly inspired by this work, we develop theoretical grounds for the use of negative control exposures in peer influence settings. 
 In order to illustrate our approach, we reconsider as running example the analysis performed by \citet{christakis2007spread} to evaluate the contagion effect of obesity using dyadic data from the Framingham Study. In the Framingham study, we consider as negative control exposure,  the alter's BMI measurement  from the subsequent visit.  %since it  likely is affected by the latent factor and it happens at the same time as the outcome and thus cannot be a cause of the latter. 
 In contrast to the IV assumption which rules out any dependence between the IV and the unmeasured factor implicated in homophily mechanism, our method requires and leverages such dependence. We provide sufficient conditions under which  our negative control exposure can be used to detect and account for homophily bias in order to recover the causal effect of primary interest. Moreover, it is worth noting that the proposed method accommodates both directional and mutual nameship in social influences.

% Such example including 

%We do not require the genes to be independent of the unmeasured factors that affects the relationship formation. On the contrary, the genes that are associated with the relationship formation could be used to induce exogeneity into the regression model for the alter's BMI. Such use of negative control exposure  variable in adjusting the homophily bias is novel to the literature. We show that our regression-based methods can adjust for the homophily bias and recover the causal effect. Moreover, our proposed method accommodates directional relationship as well as mutual relationship. 

The paper is organized as follows. In Section \ref{sec: preliminaries}, we introduce notation. We propose a general regression-based framework to adjust for homophily bias with a negative control exposure variable in Section \ref{sec: inferences}. Next, we illustrate our methods in estimating the spread of obesity in the  Framingham Offspring Study in Section \ref{sec: applications}. We conclude with a discussion in Section \ref{sec: discussion}.

%\section{Motivating Example}\label{sec: motivating_eg}
\section{Preliminaries}\label{sec: preliminaries}

%Our analysis focuses on the Framingham offspring generation since network information was not fully collected in the original cohort. 
In social-network dyadic analysis terminology, the key subjects of interest are called ``egos''  and any subjects to whom egos are linked are called ``alters.''  The roles of ego and alter are exchangeable depending on which person's outcome is of interest. To simplify the problem, we only consider data where the study population can be partitioned into pairs, or ``dyads'' in social sociology terminology. Although the approach equally applies to overlapping dyads but requires  appropriately accounting for dependence across dyads as discussed in \citet{vanderweele2012why}.  Following the notation of \citet{o2014estimating}, we use subscript 1 to denote alter and 2 to denote ego for any given dyad. We focus on the spread of a trait between two time points. That is, we take the perspective of individual 2 and the goal is to estimate the effect of individual 1's trait at baseline on the trait of individual 2 at follow-up. For example, in Framingham Offspring Study, we are interested in the effect of having an obese person as alter at baseline on ego's BMI status at a subsequent study visit. Such information is important for clinical and public health interventions \citep{christakis2007spread}. %Although we only consider the spread of individual traits in the network over one time lag, generalization of multiple time lags could be made easily.

We consider a study design where the dyads are  based on  nameship. As in Framingham Offspring Study, each study participant is required to name a single person of contact in an effort to mitigate loss to follow-up. A dyad is formed between two persons if at least one person names the second.  Let $R_{1}=1$ if  alter names ego as their contact person at baseline and otherwise $R_{1}=0$. Similarly,  let $R_{2}$ denote whether ego names alter as their contact  at baseline. We restrict nameship variables $R_{1}$ and $R_{2}$ within a dyad. Because both $R_1$ and $R_2$ are binary variables, there are four different nameship types, which we encode with $S$: (a) null naming $S=0$ if $(R_1,R_2)=(0,0)$; (b) active naming $S=1$ if $(R_1,R_2)=(0,1)$; (c) passive naming $S=2$ if $(R_1,R_2)=(1,0)$ and (d) mutual naming $S=3$ if $(R_1,R_2)=(1,1)$. Active naming indicates ego names  alter while the alter does not name the ego. Passive naming indicates alter names the ego while the ego does not name the alter. Null naming indicates neither individual names the other while mutual naming indicates both individuals name the other. Because dyad formation requires at lease one person naming another,  $S\geq1$ in the  observed sample of dyads.  

Let $Y^{b}_i$ and $Y^{1}_i$ denote the observed traits of individual $i$ at baseline and at follow-up $i=1,2$. The outcome of interest is ego's trait at follow-up, i.e., $Y^{1}_2$.  For clarity sake,  subscripts and superscripts are sometime suppressed, such as $Y=Y^{1}_2$. Let $A$ denote ego's exposure value, i.e., that is, the indicator of alter's trait at baseline. For example, in the case where obesity defines the trait of interest, $A$ is alter's obesity status, i.e., $A=1(\text{alter's BMI }
\geq30)$. Our methods apply more generally, whether $A$ is binary, continuous, polytomous or a count exposure. Let $a$ be a possible realization of $A$ (e.g., $a=1$ for obese and $a=0$ for no obese), and $Y(a)$ denote an ego's potential outcome if her exposure were hypothetically set to $a$. Throughout, we make the consistency assumption that the observed outcome is $Y=Y(a)$ almost surely, when $A=a$.% at time $t-1$. 

%Let $N_i(a)$ to be the potential negative control outcome under possible treatment $A_i=a$. Let $N_i$ denote the observed negative control outcome that is not directly influenced by exposure $A_i$, i.e., $N_i=N_i(0)=N_i(1)$. The choice of negative control outcome usually requires subject matter knowledge. However, we argue that in a longitudinal study such as Framingham Offspring Study, the outcome measure of ego at the previous time point $t-1$ serves a natural negative outcome, i.e., $N_i=Y_i^{t-1}$. The reason is that both the exposure $A$ and the negative control $N$ was measured simultaneously at time $t-1$ and thus the assumption of one does not have effect on the other is plausible. 

\begin{figure}
\caption{Causal diagram illustrating homophily bias.\label{fig: homophily_DAG}}
\centering
% \caption{Hajek 2 causal effect estimators for the rotavirus vaccine study, including single child household, 7 digits neighbor}\label{fig: dia_haj2}
\includegraphics[scale=.6]{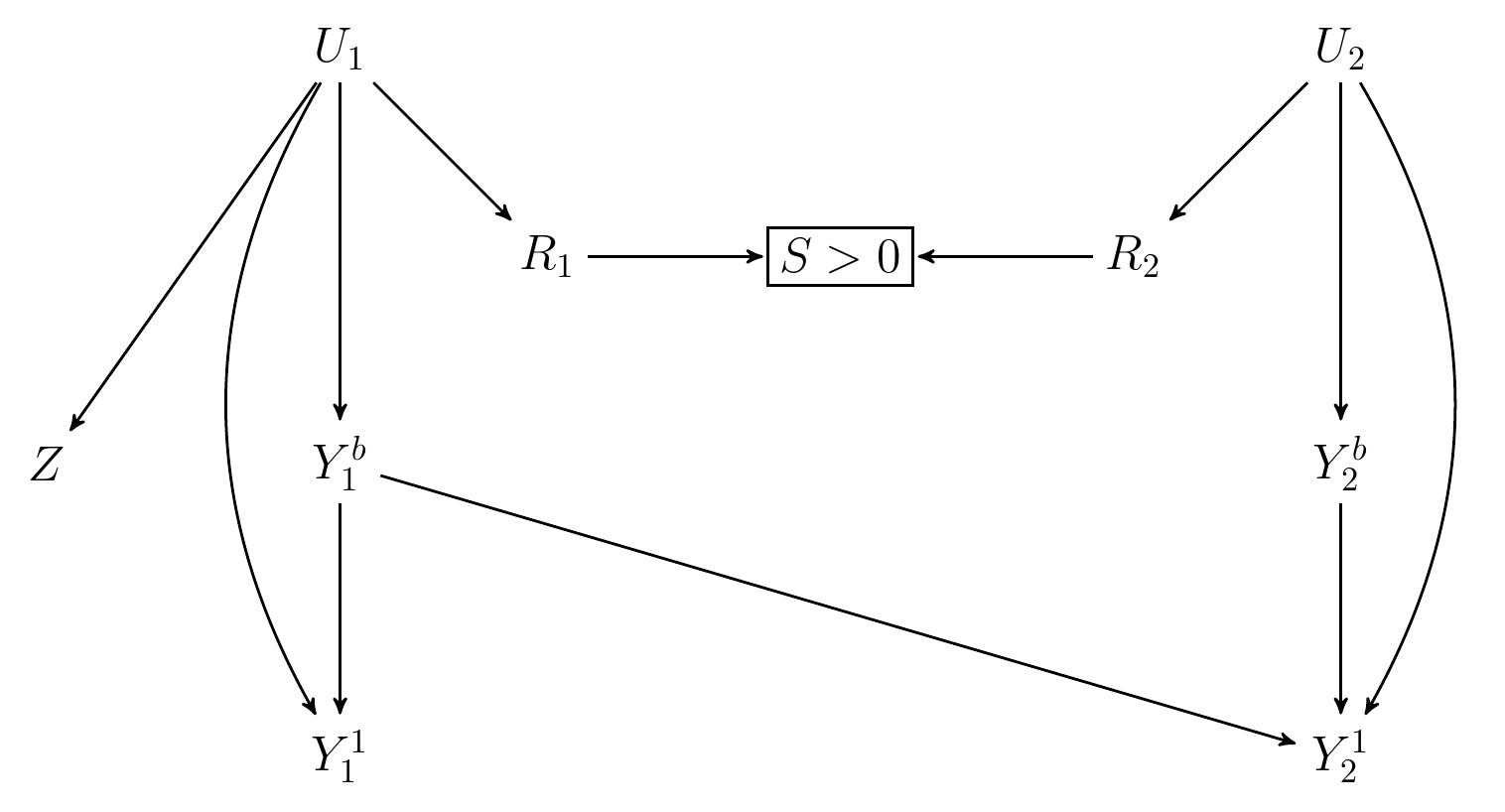}\\
\caption{The parameter of interest is the effect of the obesity status of alter (individual 1) at baseline on ego BMI  (individual 2) at  follow-up, i.e., $A=Y_1^b$, $Y=Y_2^1$. We use $Y_i^{b}$ and $Y_i^{1}$ to denote the observed weight information on individual $i$ baseline and follow-up, $U_i$ is the unmeasured factor that affects both the nameship and the weight of individual $i$, $R_{i}$ is the nameship variable for individual $i$ and $S$ is the summary of nameship type. We omit observed covariates $C_i$ for simplicity. In our empirical example, we use $Y_1^1$ as the negative control exposure $Z$.}
\end{figure}

% which includes the BMI of ego at time $t-1$ $Y^{t-1}_1$
Let $C_1$, $C_2$ denote the observed covariates at baseline for alter and ego respectively including exposure variables and let $C$ denote covariates for alter and ego excluding the exposure variable, i.e., $C=(C_1,C_2)\backslash A$. In Framingham Offspring study, we include in $C$: age, sex of both alter and ego, interaction between ego and alters' age, and ego's baseline BMI status. Let $U_1$ denote an unmeasured factor that affects not only past and current traits of the  alter ($Y^{b}_1$, $Y^{1}_1$), but also the nameship  variable $R_1$. Define $U_2$ similarly. The corresponding directed acyclic graph is given in Figure \ref{fig: homophily_DAG} (Shalizi and Thomas, 2011\nocite{shalizi2011homophily}).  The parameter of interest is $\gamma_{s,c}=E\{Y(1)-Y(0)|S=s,C=c\}$ for $s=1,2,3$, which corresponds to the average treatment effect of the alter's baseline trait on ego's trait at the follow-up visit, given that the dyad is of type $s$ and covariates $C=c$.

Because for all observed dyads,  $S\geq1$, the DAG in Figure \ref{fig: homophily_DAG} represents the conditional distribution of $(Y,A, C)$ conditional on $S\geq1$. Because $S$ is a descendant of both $U_1$ and $U_2$, in the terminology of graph theory, $S$ is called a collider\footnote{Conditioning on collider $S$ or its descendant unblocks a back-door path $A-U_1-R_1-S-R_2-U_2-Y$ (Pearl, 2009\nocite{pearl2009causality}).}
 (Pearl, 2009\nocite{pearl2009causality}, Shalizi and Thomas, 2011\nocite{shalizi2011homophily}). A direct consequence of this graphical structure is that a standard regression model for $Y$ conditional on $S$, $C$ and $A$,  which fails to condition on either $U_1$ or $U_2$ will generally be subject to collider bias so that it may reveal a non-null  association between $A$ and $Y$ even when $A$ fails to cause $Y$ and there is no unmeasured confounding of the effects of $A$ on $Y$ in the underlying population (see Figure \ref{fig: homophily_DAG}).  This specific type of collider bias is called homophily bias. Because $U_1$ and $U_2$ are unobserved and $S\geq 1$ is always conditioned on, homophily bias \citep{shalizi2011homophily} %back-door path
  cannot be accounted for without an additional assumption. Next we consider leveraging a negative control exposure to both detect and correct for collider bias. 
  
Let $Z$ denote a negative control exposure variable that satisfies the following assumptions: 

\begin{assumption}\label{assumpt_1}
$Z\nindep S|A,C$; 
\end{assumption}

%\noindent and 
%
%\begin{assumption}\label{assumpt_2}
%$Y(a,z)=Y(a)$ almost surely for all $a$ and $z$; 
%\end{assumption}
%

\begin{assumption}\label{assumpt_nodirect_cause}
$Y(a,z)=Y(a)$ almost surely;
\end{assumption}

\begin{assumption}\label{assumpt_exclusion}
$Z\indep Y(a,z)|A, C,S,U_2$,
\end{assumption}

\noindent where $\indep$ denotes independence between variables and $\nindep$ denotes dependence. Assumption \ref{assumpt_1} states that $Z$ must be associated with $S$ given $A$ and $C$. This assumption is represented in the DAG of Figure \ref{fig: homophily_DAG}, provided that the arrow between $U_1$ and $Z$ is known to be present. The assumption would also hold if $Z$ where a direct cause of $R_1$  even if $Z$ were independent of $U_1$.  %Assumption \ref{assumpt_2} states that the negative control variable $Z$ does not causally affect $Y$.  Although $Z$ does not have a causal relationship with outcome $Y$, they may still be associated due to homophily bias.
 Assumption \ref{assumpt_nodirect_cause} is a form of exclusion restriction of no direct causal effect of $Z$ on $Y$ upon setting $A$ to $a$. Assumption  \ref{assumpt_exclusion} is an assumption of no unmeasured confounding between $Z$ and $Y$ conditional on $A$, $C$, $S$, and $U_2$.
Thus, the association between $Z$ and $Y$ given $A,C,S$ can  be attributed completely to homophily bias. Hereafter, a negative control exposure for homophily bias control is a variable known to satisfy Assumptions \ref{assumpt_1}-\ref{assumpt_exclusion}. 

Furthermore, we assume that the exposure variable is not subject to unmeasured confounding given $(C,S,U_2)$ as illustrated in the DAG in Figure \ref{fig: homophily_DAG}:

\begin{assumption}\label{assumpt_ignore}
$A\indep Y(a)|C,S,U_2$.
\end{assumption}
Assumption \ref{assumpt_ignore} rules out residual confounding of the causal effect of $A$ on $Y$ upon conditioning on $C$, $U_2$  and nameship type $S$.  However, $A$ is not independent of $Y(a)$ given $C$ and $S$ only and therefore, homophily may be interpreted as inducing a form of unmeasured confounding by $U_2$ upon conditioning on $S$, even though $U_2$ is not a common cause of $A$ and $Y$ in the overall population (i.e., upon marginalizing over $S$).

%The role of a negative control variable $Z$ in the homophily problem is distinct from that of an instrumental variable in that the latter does not allow the variable to be affected by the latent factor while the former allows and leverages such relationship. However, they also share the similarity to introduce exclusion restriction into the model. 

The following two examples provide choices of negative control exposures that have been considered in social studies. 
 %In dyadic studies, traits of the alter from the subsequent visits are reasonable choices for $Z$ when the parameter of interest is the peer effect of alter's traits on ego at a subsequent visits. This is because such variables likely are affected by the latent factor and unlikely to be causes of the ego's traits at a subsequent time as the two happens simultaneously. For example, we could choose the traits or characteristics of the ego, such as health status, physical activities at a later time point as the negative control. 

\begin{example}
\citet{elwert2008wives} investigated the potential presence of homogamy bias (homophily bias due to spousal similarity) in making inference about the widowhood effect. Specifically, they proposed to use  the potential death of an ex-wife as a negative control exposure of the widowhood effect on the mortality of their ex-husband to test for 
homogamy bias. They found a significant
effect of a current wife’s death on her husband’s mortality but no significant
effect of an ex-wife’s death on her ex-husband’s mortality. These results support the existence of a causal widowhood effect, which cannot be  explained away by homogamy bias.
\end{example}

\begin{example}
\citet{cohen2009detecting} applied the regression methods in \citet{christakis2007spread} and \citet{christakis2008collective} to traits that are unlikely to be transmitted via social networks including acne, headaches, and height. They found that these traits are significantly associated among friends and thus conclude the existence of homophily bias in the network effects in the literature. Technically, these analyses may be viewed as double negative control analyses as they incorporate both negative control exposure and outcome variables \citep{miao2017invited,miao2018identifying}.
\end{example}

We reanalyze the Framingham data considered by \citet{christakis2007spread} using our proposed methodology taking as negative control exposure variable, the ego's BMI measure at follow-up $Z=Y_1^1$. Ego and alter's contemporaneous BMI measures cannot be causally related, therefore fullfilling Assumption \ref{assumpt_nodirect_cause}. Furthermore, it is clear that such a choice of $Z$ is guaranteed to satisfy Assumption \ref{assumpt_1} because any unmeasured cause of ego's baseline BMI (and S) is likely also a cause of his or hers BMI at follow-up.  In Section \ref{sec: inferences}, we provide conditions under which Assumption %s \ref{assumpt_nodirect_cause} and 
\ref{assumpt_exclusion} is also credible for this choice of negative control exposure.

\section{Regression Based Approach}\label{sec: inferences}
\subsection{Identification}

We first discuss the case where $Y$ is continuous. Suppose the data generating mechanism satisfies 

%Specifically, we consider the following regression model:

%\vspace{-5mm}
\begin{equation}\label{eq: reg_Y_general}
E(Y|S=s,A,C,Z,U_2)=U_2+b^s(A,C)+\tau^s(C),
\end{equation}

%\noindent where $b^s(A,C)$ is an unrestricted function of the exposure $A$ and covariates $C$, which may vary with different relationship type and $b^s(0,C)=0$. 
\noindent where $b^s(0,C)=0$, $b^s(A,C)$ and $\tau^s(C)$ are otherwise unrestricted. The outcome regression model \eqref{eq: reg_Y_general} assumes that the effect of $U_2$ on ego's trait does not interact with  $A$. %Model \eqref{eq: reg_Y_general} conditions on the latent factor $U_2$, thus removing all possible homophily bias. 
Under Assumptions \ref{assumpt_nodirect_cause}--\ref{assumpt_exclusion} encoded in the model, the right-hand side of Model \eqref{eq: reg_Y_general} does not depend on $Z$.  Furthermore, under  Assumption \ref{assumpt_ignore}, The conditional causal effect  of interest under Model \eqref{eq: reg_Y_general} is $E\{Y(1)-Y(0)|S=s,U_2,C,Z\}=E\{Y(1)-Y(0)|S=s,C\}=b^s(1,C)-b^s(0,C)$. For example, in Framingham Offspring Study, the parameter of interest can be interpreted as the contagion effect in nameship $s$ of alter's obesity status at  baseline on ego's BMI at the follow-up visit within levels of $C$. A detailed derivation of the causal contagion effect is given in the Appendix. The standard linear structural model is a special case corresponding to $b^s(A,C;\beta_a^s)=\beta_a^sA$, $\tau^s(C;\beta^s_c)={\beta^{s}_c}^T C$, where $T$ denotes matrix transpose. 
%While our regression method proposed below is applicable for model \eqref{eq: reg_Y_general}, f
%To simplify parameter of interest, we relegate the identification and estimation results for model \eqref{eq: reg_Y_general} to the Appendix and only base our main discussion on the following model \eqref{eq: reg_Y}. Controlling for the unobserved latent factor $U_2$, we assume
%
%
%
%
%
%\begin{equation}\label{eq: reg_Y}
%E(Y|S=s,A,C,Z,U_2;\beta^s)=\beta_0^s+\beta_a^sA+U_2+b^s(C),
%\end{equation}
%
%\noindent where $\beta^s=(\beta^s_0,\beta^s_a)$, $b^s(C)$ is an unrestricted function of covariates $C$. Model \eqref{eq: reg_Y} is a special case of model \eqref{eq: reg_Y_general} when $b^s(A,C)=\beta_a^sA+b^s(C)$. In model \eqref{eq: reg_Y}, the causal contagion effect of alter's obesity status on ego's BMI  is $\beta_a^s$.%Under model \eqref{eq: reg_Y}, the causal effect of interest $E\{Y(1)-Y(0)|S=s,C\}$ is simply $E\{Y(1)-Y(0)|S=s,C\}=\beta_a^s$ for $s=1,2,3$. Model \eqref{eq: reg_Y_general} reduces to model \eqref{eq: reg_Y} when $b^s(A,C)=\beta_a^sA+b^s(C)$. Under model \eqref{eq: reg_Y_general}, the causal effect of interest is $E\{Y(1)-Y(0)|S=s,C\}=b^s(1,C)-b^s(0,C)$. Note that our method could be easily extended to the identification and estimation of parameters in model \eqref{eq: reg_Y_general}. 

However, because $U_2$ is unobserved, an additional assumption is needed for identification. We consider the following generalized polytomous logit model for $S|A,C,Z$ and $U_2$

%\vspace{-5mm}
\begin{equation}\label{eq: reg_S}
\log \frac{\Pr(S=s|A,C,Z,U_2)}{\Pr(S=0|A,C,Z,U_2)}=\alpha^s U_2+\gamma^s(A,C,Z),
\end{equation}

\noindent where $\gamma^s(A,C,Z)=\log \Pr(S=s|A,U_2=0,C,Z)/\Pr(S=0|A,U_2=0,C,Z)$ is the baseline log odds function of $S=s$ when $U_2$ is set to its reference value 0. Equation \eqref{eq: reg_S} specifies a  log linear odds ratio association between $U_2$ and $S$ conditional on $A$, $C$ and $Z$ while leaving $\gamma^s(A,C,Z)$ unrestricted. An important example within this class of models we will primarily focus on is given by a multinomial logistic regression $\log \{\Pr(S=s|A,C,Z,U_2)/\Pr(S=0|A,C,Z,U_2)\}=\alpha^s U_2+\gamma_1^sA+\gamma_2^sC+\gamma_3^sZ$.

Additionally, we assume that in the population, $U_2$ and $(A,Z)$ are mean independent conditional on $C$:

%\vspace{-5mm}
\begin{equation}\label{eq: mean_indep_U_pop}
E(U_2|A,C,Z)=E(U_2|C).
\end{equation}

\noindent Assumption \eqref{eq: mean_indep_U_pop} is consistent with the causal diagram in Figure \ref{fig: homophily_DAG} because $U_2$ and $A,Z$ are marginally independent for any pair of individuals in the underlying population, i.e. in absence of collider bias induced by conditioning on $S$.

Finally, we assume that

%\vspace{-10mm}
\begin{equation}\label{eq: reg_U}
\Delta_s\indep (A,Z)|S,C,
\end{equation}

\noindent where $\Delta_s=U_2-E(U_2|S=s,A,C,Z)$. Assumption \eqref{eq: reg_U} states that conditional on $C$ and $S$, the association between $U_2$ and $(A,Z)$ is entirely due to a location shift. This assumption would hold if $U_2$ were normally distributed with homoscedastic error, conditional on $S=s,A,C,Z$. In principle, as apparent in proving our main results, equation (4)  only needs to hold for $S=0$, and therefore selection bias may in fact be more severe for dyads with $S\neq0$ so that association between $\Delta_s$ and $(A,C)$ may manifest itself beyond the mean in these dyads, e.g., with the shape and spread of $U_2$.  
%Actually, for the identification of causal effect in Model \eqref{eq: reg_Y_general}, we need assumption  \eqref{eq: reg_U} to hold only for the null naming relationship $s=0$, i.e., $\Delta_0\indep (A,Z)|S=0,C$.  %Assumptions \eqref{eq: mean_indep_U_pop} and \eqref{eq: reg_U} hold under the linear model $U_2=\eta_1A+\eta_2Z+\eta_3^TC+\delta^s$, where $\delta^s$ follows $N(0,\sigma^2)$ for $s=0,\ldots,3$.

Assumptions \eqref{eq: reg_Y_general}--\eqref{eq: reg_U} are not testable without an additional restriction. The following example illustrates a familiar shared random effect model under which equations \eqref{eq: reg_Y_general}--\eqref{eq: reg_U} hold.

%Intuitively, they are imposed such that the homophily bias is separated from the causal effect of interest $b^s(A,C)$. The following example provides a scenario where Assumptions \eqref{eq: reg_Y_general}--\eqref{eq: reg_U} hold under common models.

\begin{example}\label{eg: gaussian}
Suppose that $E(Y|S=s,A,C,Z,U_2)=U_2+\beta_a^s(C)A+{\beta^{s}}^T C$, \[\log \frac{\Pr(S=s|A,C,Z,U_2)}{\Pr(S=0|A,C,Z,U_2)}=\alpha^s U_2+\gamma_1^sA+{\gamma_2^s}^TC+\gamma_3^sZ\] and $U_2$ is the random effect shared between models for $Y$ and $S$ to encode a latent association between them with $U_2|S=s,A,C,Z\sim N(\eta^T C,\sigma^2)$, $s=0,\ldots,3$, then Assumptions \eqref{eq: reg_Y_general}--\eqref{eq: reg_U} hold.
\end{example}

 %The assumption \eqref{eq: reg_U} is approximately correct if the degree of homophily bias is small; however, the assumption is otherwise not testable without an additional restriction, for instance, one could test this assumption if further restrict the density of $U_2$ to belong to a certain parametric family.
We now give our main identification result under Model \eqref{eq: reg_Y_general}.

\begin{proposition}\label{prop: reg_general}
Under Model  \eqref{eq: reg_Y_general}, %assume the negative control exposure is valid, and the exposure is ignorable without the homophily, i.e.,  
 Assumptions \ref{assumpt_1}--\ref{assumpt_ignore} and equations  \eqref{eq: reg_S}--\eqref{eq: reg_U}, we have that
\begin{equation}\label{eq: reg_general}
E(Y|A,C,Z,S=s)=\sum_{\tilde s\neq s}\beta^{s\tilde s}(C)\Pr(S=\tilde s|A,C,Z)+b^s(A,C)+\bar\tau^{s}(C),
\end{equation}
\noindent  where  $\bar\tau^{s}(C)$ is an unrestricted function of $C$, $\beta^{s\tilde s}(C)=E(U_2|A,C,Z,S=s)-E(U_2|A,C,Z,S=\tilde s)$ %an explicit formula for $\beta^{s\tilde s}(C)$ is provided in the Appendix.%$\beta^{s\tilde s}(C)=E(U_2|C,S=s)-E(U_2|C,S=\tilde s)$. %Additionally, the contagion effect $E\{Y(1)-Y(0)|S=s,C\}$ is identified.
\end{proposition}

%
%
%\begin{proposition}\label{prop: reg}
%Under model \eqref{eq: reg_Y_general} and assuming \eqref{eq: reg_S}--\eqref{eq: reg_U}, we have that
%
%%\vspace{-8mm}
%\begin{equation}\label{eq: reg}
%E(Y|S=s,A,C,Z;\beta^{s*})=b^s(A,C)+\sum_{\tilde s\neq s}\beta^{s\tilde s}(C)\Pr(S=\tilde s|A,C,Z)+b^{s\ast}(C),
%\end{equation}

%%\vspace{-5mm}
%\noindent  where $\beta^{s*}=(\beta^s_a,\beta^{s\tilde s}(C))$ and $b^{s\ast}(C)$ is an unrestricted function of $C$.
%%, the contrast $\beta^{s\tilde s}(C)$ does not depend on $A$.
%\end{proposition}

%From Proposition \ref{prop: reg_general}, we cannot identify $b^s(A,C)$ or $\bar\tau^{s}(C)$, however, we could identify $b^{*s}(A,C)=b^s(A,C)+\bar\tau^{s}(C)$. This is sufficient to identify the causal effect since $E\{Y(1)-Y(0)|S=s,C\}=b^s(1,C)-b^s(0,C)=b^{*s}(1,C)-b^{*s}(0,C)$.
%
%
%**********
We provide a detailed proof in the Appendix. %and give description of what result looks like in Gaussian case. %and give a brief intuitive illustration here. %Briefly, under the assumptions of Proposition \ref{prop: reg_general}, we obtain a regression model \eqref{eq: reg_general}, which facilitate identification and estimation. 
Comparing \eqref{eq: reg_general} with \eqref{eq: reg_Y_general}, we note that the left hand-side of  \eqref{eq: reg_general} is by iterated expectation equal to $E\{E(Y|A,C,U_2,S=s)|A,C,Z,S=s\}=E(U_2|A,C,S=s,Z)+b^s(A,C)+\tau^s(C)$, and therefore the proof of Proposition \ref{prop: reg_general} hinges on establishing that under our assumptions $E(U_2|A,C,S=s,Z) =\sum_{\tilde s\neq s}\beta^{s\tilde s}(C)\Pr(S=\tilde s|A,C,Z)+\bar\tau^s(C)-\tau^s(C)$. Equation \eqref{eq: reg_general} highlights the important role of the negative control variable $Z$ which appears on the right hand side of the equation only through its association with $S$ in $\Pr(S=\tilde s|A,C,Z)$.  Note that equation \eqref{eq: reg_general} would continue to hold even if $Z$ were not conditioned on (or the edge from $Z$ to $U_1$ were removed in Figure \ref{fig: homophily_DAG}, such that $Z$ were independent of $U_1$ given $A, C,S$), with $\Pr(S=\tilde s|A,C)$ in for $\Pr(S=\tilde s|A,C,Z)$, in which case it would generally not be possible to tease apart this latter term which captures selection bias from structural part of the equation $b^s(A,C)$ as both are unrestricted function of $(A,C)$, thus rendering the causal effect non-identified. Identification of the causal contagion effect now depends on identification of $\Pr(S=\tilde s|A,C,Z)$ given dyadic study design. Below, we provide sufficient conditions under which such identification is possible.

According to Proposition \ref{prop: reg_general},  the coefficient $\beta^{s\tilde s}(C)=E(U_2|A,C,Z,S=s)-E(U_2|A,C,Z,S=\tilde s)$. %, which does not depend on $A$ and $Z$ due to assumption \eqref{eq: reg_U}.
 Hence, $\beta^{s\tilde s}(C)$ encodes the association between $S$ and $U_2$ and therefore is zero if either $U_2$ does not predict $S$, i.e., $\alpha^s$ is the same for all $s$, or if $U_2$ is degenerate in the sense that it does not predict $Y$.  In the Gaussian case of Example \ref{eg: gaussian}, we show in the Appendix that $\beta^{s\tilde s}(C)=\sigma^2(\alpha^s-\alpha^{\tilde s})$ making explicit the aforementioned interpretation. An important advantage of the proposed approach is that it provides a framework to formally test the null hypothesis of no homophily bias as a test of the null hypothesis that 
 $\beta^{s\tilde s}=0$ for all $s,\tilde s$.  

Proposition \ref{prop: reg_general}  assumes the identity link function for the outcome model.  Similar results can be obtained for a multiplicative model (i.e. log link)  which is more appropriate for binary or count outcomes. For  instance, when the response is binary, the following conditional causal risk ratio may be of interest  $P\{Y(1)=1|S,C\}/P\{Y(0)=1|S,C\}$ for $s=1,2,3$. To ground ideas, suppose that %conditional on the unobserved factor $U_2$ 

\begin{equation}\label{eq: reg_Y_bin}
\log E(Y|S=s,A,C,Z,U_2;\beta^s)=U_2+b^s(A,C)+\bar\tau^{s}(C).
\end{equation}
\noindent Because $U_2$ is conditioned on in \eqref{eq: reg_Y_bin}, suppose Assumption \ref{assumpt_ignore} holds, $\exp\{b^s(1,C)-b^{s}(0,C)\}$ can be interpreted as the causal contagion effect of alter on ego on the multiplicative scale, e.g. on the risk ratio scale for binary $Y$. A similar effect can be defined when the treatment $A$ is continuous. We have the following result for the multiplicative model, the proof of which is given in the Appendix. With a slight abuse of notation, we use the same notation for parameters as in the case of the additive model. 

\begin{proposition}\label{prop: reg_bin}
Under Model  \eqref{eq: reg_Y_bin}, Assumptions \ref{assumpt_1}--\ref{assumpt_ignore} and equations \eqref{eq: reg_S}--\eqref{eq: reg_U}, we have 

%\vspace{-8mm}
\begin{equation}\label{eq: reg_bin}
\log E(Y|S=s,A,C,Z;\beta^{s\tilde s}(C))=\sum_{\tilde s\neq s}\beta^{s\tilde s}(C)\Pr(S=\tilde s|A,C,Z)+b^s(A,C)+\bar\tau^{s}(C),
\end{equation}
\noindent  where  $\bar\tau^{s}(C)$ is an unrestricted function of $C$. %Additionally, the contagion effect on the multiplicative scale $P\{Y(1)=1|S,C\}/P\{Y(0)=1|S,C\}$ is identified.
% the contrast $\beta^{s\tilde s}(C)$ does not depend on $A$.
\end{proposition}

% We need to say something about identification of S mechanism before estimation and inference. see suggestion below. 

Propositions \ref{prop: reg_general} and \ref{prop: reg_bin} are only useful to the extent that one can identify the selection mechanism $Pr(S|A,C,Z)$ from observed dyadic sample. Because the sample implicitly conditions on $S\geq 1$, nonparametric identification is in general not an option, and therefore one must impose a restriction in order to make progress. In this vein, we propose to posit a model of form $\Pr(S|A,C,Z ;\theta)$ with finite dimensional unknown parameter $\theta$. %In order to motivate the approach 

\subsection{Estimation and Inference}

%For estimation, Model \eqref{eq: reg_general} is quite general. 

%We only illustrate the estimation procedure with Model \eqref{eq: reg_general} because that of Model \eqref{eq: reg_bin} is similar.  Semiparametric or nonparametric methods, such as kernel smoothing, splines, polynomials or wavelets could be used to model the covariates. The parameters in Model  \eqref{eq: reg_general} can be estimated in a two-stage fashion. At the first stage, the estimated propensity score could be obtained, for example, by the maximal likelihood estimation (MLE) if a parametric propensity score model is assumed. The estimated propensity score is then plugged in  \eqref{eq: reg_general} to obtain estimates for the coefficients in the outcome regression model. 

 Specifically, we make the following assumption to simplify the estimation procedure.
 
 \begin{assumption}\label{assump: R_indep}
 %Assume that $R_1$ and $R_2$ are conditionally independent given $(A,C,Z)$, i.e., 
 $R_1\indep R_2|(A,C,Z)$. 
 \end{assumption}
 
\noindent Under Assumption \ref{assump: R_indep}, we may specify the following parametric model  \[\Pr(S=\tilde s|A,C,Z;\theta)=\Pr(R_1|C_1,Z;\theta_1)\Pr(R_2|C_2;\theta_2),\] where $\theta=(\theta_1^T,\theta_2^T)^T$, $R_1$ is independent of $C_2$ given $C_1,Z$ and $R_2$ is independent of $C_1, Z$ given $C_2$, $\Pr(R_i=1|C_i,Z_i;\theta_i)$ follows a logistic regression with mean $\exp(\theta_i^T\tilde C_i)/\{1+\exp(\theta_i^T\tilde C_i)\}$, where $\tilde C_1=(C_1,Z)$ and $\tilde C_2=C_2$. Then, we have

%\vspace{-.5cm}
\begin{equation*}
\Pr(R_1,R_2|A,C,Z;\theta)=\frac{\exp(\theta_1^T\tilde C_1R_1+\theta_2^T\tilde C_2R_2)}{\{1+\exp(\theta_1^T\tilde C_1)\}\{1+\exp(\theta_2^T\tilde C_2)\}}.
\end{equation*} 

\noindent It follows that  %By the relationship between $S$ and $(R_1,R_2)$, we can  derive the probability $\Pr(S=s|A,C,Z)$. For example, we have

%\vspace{-.5cm}
\begin{equation*}
\Pr(S|A,C,Z;\theta)=\frac{1(S=1)\exp(\theta_2^T\tilde C_2)+1(S=2)\exp(\theta_1^T\tilde C_1)+1(S=3)\exp(\theta_1^T\tilde C_1+\theta_2^T\tilde C_2)}{\{1+\exp(\theta_1^T\tilde C_1)\}\{1+\exp(\theta_2^T\tilde C_2)\}}.
\end{equation*} 
%
%%\vspace{-1.5cm}
%\begin{equation*}
%\Pr(S=2|A,C;\theta)=\Pr(R_1=1,R_2=0|A,C;\theta)=\frac{\exp(\theta^T\tilde C_1)}{\{1+\exp(\theta^T\tilde C_1)\}\{1+\exp(\theta^T\tilde C_2)\}},
%\end{equation*} 
%
%\noindent and
%
%%\vspace{-1.5cm}
%\begin{equation*}
%\Pr(S=3|A,C;\theta)=\Pr(R_1=1,R_2=1|A,C;\theta)=\frac{\exp\{\theta^T(\tilde C_1+\tilde C_2)\}}{\{1+\exp(\theta^T\tilde C_1)\}\{1+\exp(\theta^T\tilde C_2)\}}.
%\end{equation*} 
%

\noindent Because the observed sample space conditions on $R_1+R_2\geq 1$, the observed likelihood function for  nameship mechanism for a given dyad is given by 

%\vspace{-.5cm}
\begin{equation*}
\Pr(R_1,R_2|R_1+R_2\geq 1,A, C,Z;\theta)=\frac{\exp(\theta_1^T\tilde C_1R_1+\theta_2^T\tilde C_2R_2)}{\exp(\theta_1^T\tilde C_1)+\exp(\theta_2^T\tilde C_2)+\exp\{\theta_1^T\tilde C_1+\theta_2^T\tilde C_2\}},
\end{equation*}

\noindent the conditional log-likelihood is, 

%%\vspace{-1.5cm}
\begin{eqnarray*}
l(\theta)&=&\sum_{j=1}^J\log\Pr(S_j|S_j\geq1,\tilde C_{j1},\tilde C_{j2};\theta)\\
&=&\sum_{j=1}^J\theta_1^T\tilde C_{j1}1(S_j=1)+\theta_2^T\tilde C_{j2}1(S_j=2)+(\theta_1^T\tilde C_{j1}+\theta_2^T\tilde C_{j2})1(S_j=3)\\
&&-\log\{\exp(\theta_1^T\tilde C_{j1})+\exp(\theta_2^T\tilde C_{j2})+\exp(\theta_1^T\tilde C_{j1}+\theta_2\tilde C_{j2})\},
\end{eqnarray*}

\noindent where $j=1,\ldots,J$ is the index for dyad and $J$ is the total number of dyads in the study. The  maximum likelihood estimator $\hat\theta$ for $\theta$ is defined as $\hat\theta=\text{argmax}_{\theta} l(\theta)$. Once $\hat\theta$ has been obtained, we proceed by fitting Model \eqref{eq: reg_general} with the plug-in estimate $\Pr(S|A,C,Z;\hat\theta)$ to obtain the least square estimates for parameters in the model of $Y$ on regressors $\{b^s(A,C),\bar\tau^{s}(C),\Pr(S=\tilde s|A,C,Z):\tilde s\neq s\}$. The two-step estimation procedure is summarized in Algorithm \ref{alg: twostep}.

\begin{algorithm}[ht]
%>- from algorithm package
\caption{Two step estimation procedure with a negative control variable}\label{alg: twostep}
\begin{algorithmic}
%\Procedure{Algorithm for the two step estimation procedure}{}
\State Step 1. Choose a negative control variable $Z$ that satisfies Assumptions \ref{assumpt_1}--\ref{assumpt_exclusion}. Specify the nameship model $\Pr(S|A,C,Z;\theta)$. Obtain an estimate $\hat\theta$ of $\theta$ by fitting $\Pr(S|S\geq1,A,C,Z;\theta)$ to the data.
\State Step 2. Estimate coefficients in the additive Model \eqref{eq: reg_general} (or the multiplicative Model \eqref{eq: reg_bin}) using the estimated $\Pr(S|A,C,Z;\hat\theta)$.
%\EndProcedure
\end{algorithmic}

\end{algorithm}

One could in principle relax Assumption \ref{assump: R_indep} by allowing dependence between $R_1$ and $R_2$. A natural approach could involve specifying a random effects model $\Pr(R_1,R_2|A,C,Z)=\int_b\Pr(R_1|C_1,Z,b) \Pr(R_2|C_2,b)f(b)db$ where $f(b)$ follows a normal distribution with mean zero variance $\sigma_b^2$ and $\Pr(R_1|C_1,Z,b)$ and $\Pr(R_2|C_2,b)$ are logistic regressions with random intercept $b$. All parameters could be estimated by maximizing observed data likelihood as described above, which in this case would involve numerical integration to evaluate the likelihood contribution of each dyad. 

%Note the conditional independence between $R_1$ and $R_2$ is not needed for either identification or estimation purpose. We imposed this independence model  because in the population where the homophily does not exist,  this is a reasonable and convenient assumption. Alternatively, one could assume $R_1$ and $R_2$ are independent given some latent factor, which is more general.

 To derive the asymptotic distribution of the contagion effect estimator, we assume dyads are non-overlapping and people from different dyads are independent. Let $\rho=(\theta,\beta)$ denote the vector of the parameters in the nameship mechanism and the outcome regression.  Let $G_{\theta}(O_i;\theta)$ and $G_{\beta}^s(O_i;\rho)$ denote the estimating functions corresponding to $\hat \theta$ and $\hat \beta$ in nameship type $s$, such that $\hat\rho=\{\hat\theta,\hat\beta\}$ is the solution to the vector equation $\sum_{i=1}^n G^s(O_i;\rho)=0$, where $G^{s}(O;\rho)=\{G_{\theta}(O;\theta),$ $G_{\beta}^s(O;\rho)\}^T$.  Let $A^{\otimes 2}=A\otimes A^T$ denote the Kronecker product of $A$ and $A^T$ and let $\xrightarrow{d}$ denote convergence in distribution. The following proposition gives the asymptotic distribution of $\hat\rho$. A proof can be directly obtained from standard estimating equation theory \citep{stefanski2002calculus}. The resulting variance estimator with plugin parameter estimates is typically known as the sandwich estimator.

%, that is, $\beta^{\ast}=\beta_0$ or $\gamma^{\ast}=\gamma_0$

\begin{proposition}\label{prop: sandwich}
Under Model  \eqref{eq: reg_Y_general}, suppose that Assumptions \ref{assumpt_1}--\ref{assumpt_ignore} hold and that modeling Assumptions \eqref{eq: reg_S}--\eqref{eq: reg_U} hold,  then $n^{1/2}(\hat\rho-\rho)\xrightarrow{d}N(0,\Sigma_{\rho}^{s})$ as $n \to \infty$, where
 $
\Sigma^{s}_{\rho} =U^{-1} V U^{-T},$
$U=-E\{\partial G^s(O_i;\rho)/\partial^T \rho\}$, $V=E\{G^s(O_i;\rho)^{\otimes 2}\}$.
\end{proposition}

Similar to the additive model, the multiplicative model \eqref{eq: reg_bin} only involves observed variables, and thus it could be fit to the data. We also carry out estimation in a two-step fashion: the nameship mechanism is estimated at the first stage and the estimated propensity is used at the second stage to obtain parameter estimates in the regression model. Asymptotic distribution of the parameter estimates under model  \eqref{eq: reg_bin} can be obtained similar to that in Proposition \ref{prop: sandwich}.

\section{Framingham Offspring Study}\label{sec: applications}

The Framingham Offspring Study was initiated in 1971 and the study population consists of most of the offsprings of the original Framingham Heart Study cohort and the spouses of the offsprings. Clinical exams were offered every four years. During each clinical exam, the participants underwent a detailed examination including physical examination, medical history, laboratory testing, and electrocardiogram. At the end of each exam, each participant was asked to name a single friend, sibling or spouse, which was likely to be the one with the most influence. The original purpose of the naming process was to record a person of contact, but such information also revealed network ties and thus has been used to assess the social influence \citep{christakis2007spread,o2014estimating}. Among the network ties provided, approximately 50\% of the nominated friend contacts were also participants in the FHS and thus they had the same information, including BMI collected. Most spouses of FHS participants were also FHS participants. 

Therefore, by design, the Framingham Offspring Study population could be partitioned into dyads. We estimated our model with unique dyads of spousal and nearly disjoint friendship. Occasional overlap of dyads when the same person was named by multiple individuals was ignored similar to \citet{o2014estimating}. Because later visits suffered from severely low attenuation rate, we focused on the spread of obesity between baseline and the first follow-up.
%%\vspace{-5mm}

% 3070 distinct individuals
We carried out a peer effect analysis for 4849 distinct dyads for which alters are spouses, siblings, or friends of egos. The status of ego and alter was randomly assigned. In principle, one can use both assignments in single analysis, however, that required clustering analysis at the level of dayad to account for correlation within dyad. For the purpose of illustration, we considered a single contribution per dyad. Obesity status was defined as a binary variable that takes value 1 if BMI is over 30, and 0 if otherwise. Let $A=1(Y^{b}_{1}>30)$ denote the exposure of ego, that is, the indicator of alter being obese at baseline.  We were interested in the causal effect of alter's obesity status at baseline on the ego's BMI at  follow-up. Covariates $C$ included sex (1 for Female and 0 for Male),  age of both ego and alter, two-way interactions between ages of ego and alter, and ego BMI at baseline. Ages were mostly between 19 to 52 (5\% and 95\% quantile respectively). We mean centered age for both ego and alter for numerical stability. 

We first carried out a standard regression-based analysis which did not adjust for the potential homophily bias. More specifically, we first fitted a naive model without distinction among different nameships $E(Y|S=s,A,C;\beta_0,\beta_a,\beta_c)=\beta_0+\beta_aA+{\beta_c}^TC$ to the data. Results are given in Table \ref{tb: naive_all_relation}. Ego BMI at baseline was significantly associated with ego BMI at the follow-up. Adjusting for ego and alter's gender and age, alter's obesity status had a significant positively association with the ego's BMI at follow-up ($\hat\beta_a=0.24$, with standard error $0.10$ and $p$-value 0.01). This effect was subject to homophily bias. Next, we fitted a naive model  stratifying by different  nameship types, i.e., we fitted  $E(Y|S=s,A,C;\beta_0^s,\beta_a^s,\beta_c^{s})=\beta_0^{s}+\beta_a^sA+{\beta_c^{s}}^TC$ to the data. Results are given in Table \ref{tb: friendNR_naive}. %The only significant predictor for ego's BMI at the first follow-up across all three relationships was ego's BMI at baseline. The ego and alter's gender and age were significant in predicting the ego's BMI in some relationship but not the others. %For example, the gender of ego is significant for their BMI in a mutual relationship but not in one-sided relationships. %It is counterintuitive that female is expected to have significantly heavier BMI than male and mid-aged people tend to have lighter BMI than younger ones in some relationships. 
%The 
Alter's obesity status at baseline had a significant positive association on ego's current BMI in a mutual nameship ($\hat\beta_a^3=0.33$ with standard error 0.13 and $p$-value 0.01). Although this model is more informative than the naive model which does not condition on nameship type, such an effect still may not have causal interpretation due to possible homophily bias.

Next, we carried out a negative control regression adjustment for homophily bias. We selected alter's weight at follow-up as a  negative control variable, i.e., $Z=Y_1^{1}$. Alter's follow-up weight is an appropriate choice of negative control exposure because it cannot be causally related to ego's contemporaneous weight, therefore satisfying Assumptions \ref{assumpt_nodirect_cause}--\ref{assumpt_exclusion}. Such assumptions presume absence of any feedback in alter and ego weight change between baseline and follow-up, which is certainly expected under the sharp null of no contagion effect of weight, but may be violated under the alternative, as discussed in conclusion.  Because $U_1$ is associated with ego's baseline weight, it may be reasonable to expect that it would also be associated with ego's weight at follow-up Z, therefore fulfilling assumption \ref{assumpt_1}.      The parameter estimates of the nameship process are given in Table \ref{tb: propen_friendNR_new}.  Negative control variable, the alter BMI at follow-up, was significantly associated with nameship process. The estimated nameship mechanisms were then included as predictors in the outcome regression model $E(Y|A,C,Z,S=s;\beta_0^s,\beta_a^s,\beta_c^{s},\beta^{s\tilde s})=\beta_0^s+\sum_{\tilde s\neq s}\beta^{s\tilde s}\Pr(S=\tilde s|A,C,Z)+\beta_a^sA+{\beta_c^{s}}^TC$ under an assumption that $\beta^{s\tilde s} (C)$ does not depend on $C$. Outcome regression model estimates were given in Table \ref{tb: friendNR_2stage_new_sandwich}. Standard errors were estimated following Proposition \ref{prop: sandwich}. Therefore, uncertainty from both stages of estimation is reflected in both estimated standard errors and $p$-values.  %Ego's weight at baseline was still significant in predicting current weight across all relationships. %The ego's gender is not significant in the prediction of the BMI while adjusting for other factors. The BMI increases as one ages from early twenties to fifties ($\hat\beta_{c,{\text{age}}}^{s}$ ranges from 0.63 to 1.01 with $p\leq0.02$ for $s=1,2,3$). The alters' sex and age are both significant in predicting the ego's BMI for all relationships. 
Our analysis provides formal evidence that homophily bias may be operating in these data. Specifically, a subset of homophily coefficients $\beta^{s\tilde s}$ were marginally statistically significant  (for example, $\hat\beta^{12}=-20.67$ with standard error 10.57 and  $p$-value 0.05) indicating at least part of the association between ego and alter's weight within each dyad was indeed subject to homophily bias and therefore not causal.  Alter's obesity status at baseline had a  positive association with ego's BMI at the follow-up for passive-nameship and mutual nameship but not for active-nameship. However, no contagion effect remained statistically significant upon accounting for homophily bias. %Due to the fact that the previous analyses in the literature do not distinguish among relationship types, our results are not directly comparable with the existing results. %According to our analyses, while there is significant evidence of selection bias due to homophily, the magnitude of the bias itself does not appear to be substantial, as the magnitude of the effect of alter being obese at baseline on ego's BMI at the first visit is comparable in both standard analysis and the proposed method.

%That is, the obesity can spread in mutual relationship but not in one-sided relationship. Although the estimates of the contagion effect from regression method with negative control exposure  variable happen to align with naive analysis, the parameter in the naive analysis does not have a causal interpretation due to homophily bias while the contagion effect in model \eqref{eq: reg} has. Additionally,  the estimates for other parameters are also different in naive analysis  compared with regression methods, while the latters are more reasonable.%Although this result align with that in the naive analysis, the result of the regression method has causal interpretation. 

 %In the homophily problem, we have $NE=Y^{t}_j$, the outcome of an alter $j$ at time $t$ can served as a natural negative control exposure for the effect of $Y^{t-1}_j$ on $Y^{t}_i$. Also, note $NO=Y^{t-1}_i$ can be served as a natural negative control outcome.
 
% We also carried out a sensitivity analysis to classify spouses as a mutual relationship while keeping friends as a directional relationship. The results are similar to what we show here.

%\vspace{-7mm}
\section{Discussion}\label{sec: discussion}
In this paper, we have proposed a simple regression-based adjustment for homophily bias with a negative control exposure variable $Z$. The unmeasured variables $U_1$ and $U_2$ could in principal also directly affect $R_2$ and $R_1$ respectively, in which case, under our negative control assumptions the proposed approach may still apply.

\begin{figure}
\caption{Causal diagram illustrating homophily bias for multiple time points.\label{fig: homophily_DAG_multiple_t}}
\centering
% \caption{Hajek 2 causal effect estimators for the rotavirus vaccine study, including single child household, 7 digits neighbor}\label{fig: dia_haj2}
\includegraphics[scale=.6]{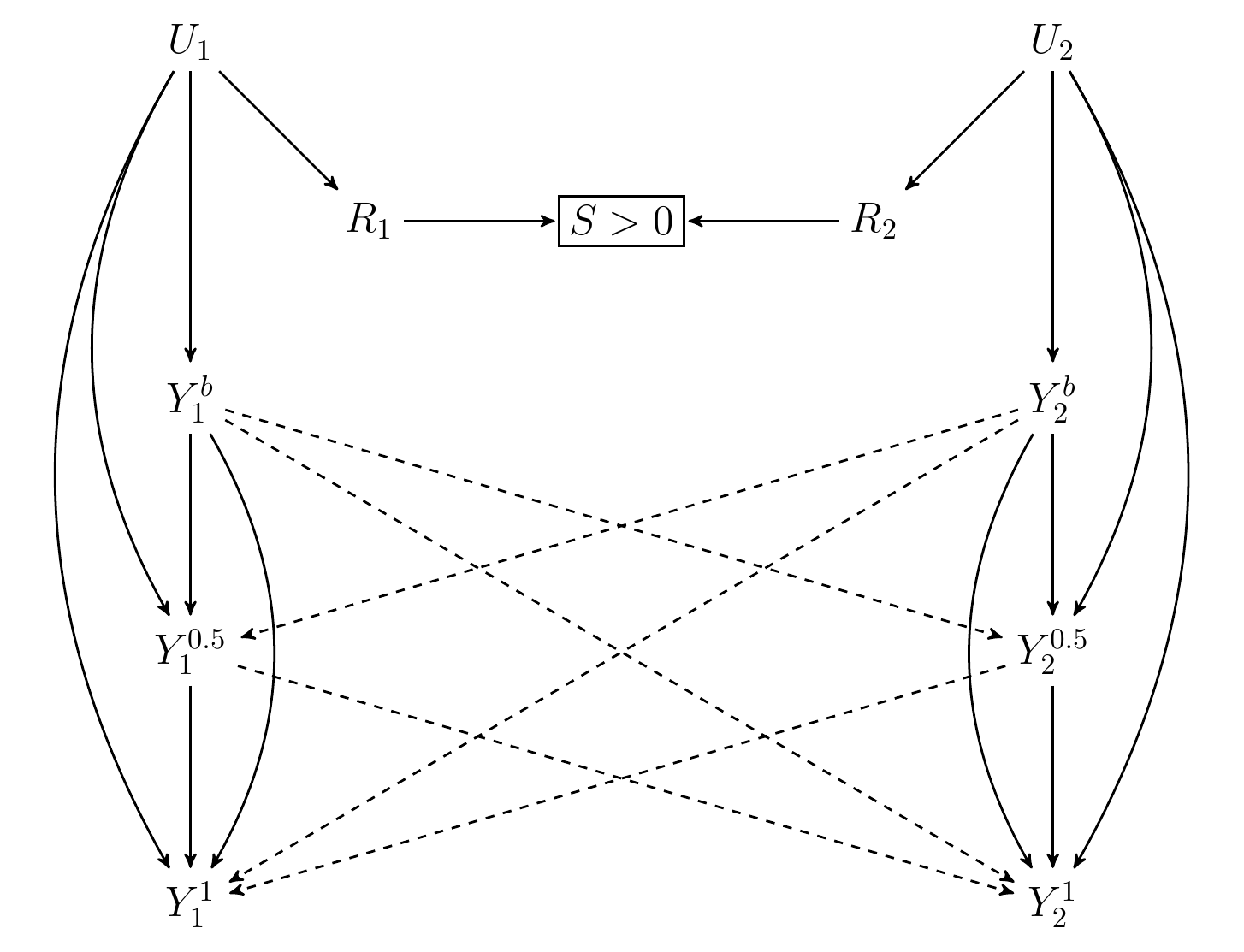}\\
\caption{The parameter of interest is the effect of the obesity status of alter (individual 1) at baseline ($A=1(Y_1^b>30)$) on ego BMI (individual 2) at time 1 ($Y_2^1$). We use $Y_i^{0.5}$ to denote the observed weight information on individual $i$ at a time point between baseline and follow-up. The dashed line denotes causal effects between individuals. We take $Z=Y_1^1$ as the negative control exposure variable.}
\end{figure}

A reviewer noted that our choice of negative control exposure in Framingham application, ego BMI  at follow-up is only applicable as a negative control variable if contagion only occurs at discrete times which are directly observed, i.e. ruling out feedback effects alluded to in Section \ref{sec: applications}.  To illustrate this, consider a situation where there is an intermediate time $t=0.5$ in between baseline and follow-up (shown in Figure \ref{fig: homophily_DAG_multiple_t}). Ego and alter  BMI can affect the other person's BMI at a follow-up visit. The dashed line denotes effects between individuals. Although alter BMI at follow-up is unlikely to have a direct causal effect on ego BMI at follow-up, they are both confounded by ego BMI at the intermediate time, $Y_1^{0.5}$. Such confounding could potentially invalidate the negative control assumption \ref{assumpt_exclusion}. %This can be circumvented by selecting a negative control variable that is collected at baseline and is known to be unlikely to have a direct causal effect of ego's subsequent BMI. For example, traits of ego such as height, acne, headache may be considered. 
This point has also been suggested in \citet{ogburn2014causal}: estimation of contagion effects at multiple time points may be complicated by the feedback issue as the entire  evolution history need to be considered.  The problem of potential uncontrolled confounding may also persist when we have multiple time points as compared with continuous time points. Because the Framingham Offspring Study follow-up was at 4 years post baseline, it is possible that causal contagion effects exist at some intermediate time between the two visits. The assumption of no unmeasured intermediate time with contagion effects is more plausible in the setting where individuals only interact during visits not in between, e.g., patients usually interact with their doctors at clinic visits. 
It is still notable as suggested in Section \ref{sec: applications}  that such complication will not occur even in Framingham Offspring Study under the sharp null hypothesis of no contagion effect, in which case, our approach would provide a valid test of the sharp null hypothesis of no contagion within 4 year window between baseline and follow-up.
%However, our method still serves as a valid method for testing the contagion under the sharp null that there are no contagion effects between individuals at any time point.

\section*{Acknowledgement}
The authors would like to give special thanks to Prof. O'Malley for insightful discussion and his patience and tremendous help on the data analysis section. The authors also thank the editor, associate editor and two reviewers for their insightful comments and helpful suggestions. Lan Liu's research is supported by NSF DMS 1916013. %\vspace*{-8pt}

%  If your paper refers to supplementary web material, then you MUST
%  include this section!!  See Instructions for Authors at the journal
%  website http://www.biometrics.tibs.org

%\section*{Supplementary Materials}
%
%Web Appendix A, referenced in Section~\ref{s:model}, is available with
%this paper at the Biometrics website on Wiley Online
%Library.%\vspace*{-8pt}

%  Here, we create the bibliographic entries manually, following the
%  journal style.  If you use this method or use natbib, PLEASE PAY
%  CAREFUL ATTENTION TO THE BIBLIOGRAPHIC STYLE IN A RECENT ISSUE OF
%  THE JOURNAL AND FOLLOW IT!  Failure to follow stylistic conventions
%  just lengthens the time spend copyediting your paper and hence its
%  position in the publication queue should it be accepted.

%  We greatly prefer that you incorporate the references for your
%  article into the body of the article as we have done here 
%  (you can use natbib or not as you choose) than use BiBTeX,
%  so that your article is self-contained in one file.
%  If you do use BiBTeX, please use the .bst file that comes with 
%  the distribution.  In this case, replace the thebibliography
%  environment below by 
%
 % \bibliographystyle{biom} 
  
 \bibliographystyle{chicago}
 \bibliography{/home/lan/Desktop/research/bibliography/mybib}

%\begin{thebibliography}{}
%
%\bibitem{ } Cox, D. R. (1972). Regression models and life tables (with
%discussion).  \textit{Journal of the Royal Statistical Society, Series B}
%\textbf{34,} 187--200.
%
%\bibitem{ }  Hastie, T., Tibshirani, R., and Friedman, J. (2001). \textit{The 
%Elements of Statistical Learning: Data Mining, Inference, and Prediction}.
%New York: Springer.
%
%\end{thebibliography}

\newpage
\appendix

%  To get the journal style of heading for an appendix, mimic the following.

\section{}
\subsection*{Proof of Proposition \ref{prop: reg_general}}

%
%model \eqref{eq: reg} with such covariate $Z$ can be written as 
%
%%%\vspace{-8mm}
%\begin{equation}\label{eq: reg_IV}
%E(Y|A,C,Z,S=s)=\beta_0^{s\ast}+\beta_a^sA+\sum_{\tilde s\neq s}\beta^{s\tilde s}(C)\Pr(S=\tilde s|A,C,Z)+b^{s\ast}(C),
%\end{equation}
%
%
%
%and model \eqref{eq: reg_bin} with $Z$ could be expressed in a similar fashion. 
%
%
%Let . 
%%%\vspace{-10mm}
%
%
%

%\textbf{Proof of Proposition \ref{prop: reg_general}:}
Under Model \eqref{eq: reg_Y_general}, we have 

%\vspace{-16mm}
\begin{eqnarray*}
&&E(Y|S=s,A,C,Z)\\
&=&E(U_2|S=s,A,C,Z)+b^s(A,C)+\tau^s(C)\\
&=&E(U_2|A,C,Z)+\sum_{\tilde s\neq s}\{E(U_2|A,C,Z,S=s)-E(U_2|A,C,Z,S=\tilde s)\}\Pr(S=\tilde s|A,C,Z)\\&&+b^s(A,C)+\tau^s(C)\\
&=&E(U_2|C)+\sum_{\tilde s\neq s}\{E(U_2|A,C,Z,S=s)-E(U_2|A,C,Z,S=\tilde s)\}\Pr(S=\tilde s|A,C,Z)\\&&+b^s(A,C)+\tau^s(C),
\end{eqnarray*}

\noindent where the last equation follows from assumption \eqref{eq: mean_indep_U_pop}. Because by assumption \eqref{eq: reg_S},

%\vspace{-12mm}
\begin{eqnarray*}
\exp(\alpha^sU_2)&=&\frac{\Pr(S=s|A,C,Z,U_2)}{\Pr(S=0|A,C,Z,U_2)}\frac{\Pr(S=0|A,C,Z,U_2=0)}{\Pr(S=s|A,C,Z,U_2=0)}\\
&=&\frac{f(U_2|A,C,Z,S=s)}{f(U_2|A,C,Z,S=0)}\frac{\Pr(U_2=0|A,C,Z,S=0)}{\Pr(U_2=0|A,C,Z,S=s)},
\end{eqnarray*}

%\vspace{-8mm}
\noindent we have 

%%\vspace{-13mm}
%\begin{eqnarray*}
%&&\frac{E\{U_2\exp(\alpha^sU_2)|A,C,Z,S=0\}}{E\{\exp(\alpha^sU_2)|A,C,Z,S=0\}}\\
%&=&E\{U_2\frac{f(U_2|A,C,Z,S=s)}{f(U_2|A,C,Z,S=0)}\frac{\Pr(U_2=0|A,C,Z,S=0)}{\Pr(U_2=0|A,C,Z,S=s)}|A,C,Z,S=0\}/\\
%&&E\{\frac{f(U_2|A,C,Z,S=s)}{f(U_2|A,C,Z,S=0)}\frac{\Pr(U_2=0|A,C,Z,S=0)}{\Pr(U_2=0|A,C,Z,S=s)}|A,C,Z,S=0\}\\
%&=&E(U_2|A,C,Z,S=s)/1\\
%&=&E(U_2|A,C,Z,S=s).
%\end{eqnarray*}
%
%%\vspace{-4mm}
%\noindent Thus,
%\vspace{-6mm}
\begin{eqnarray*}
E(U_2|A,C,Z,S=s)&=&\frac{E\{U_2\exp(\alpha^sU_2)|A,C,Z,S=0\}}{E\{\exp(\alpha^sU_2)|A,C,Z,S=0\}}\\
&=&\frac{\partial}{\partial \alpha^s}\log E\{\exp(\alpha^sU_2)|A,C,Z,S=0\}\\
&=&\frac{\partial}{\partial \alpha^s}\log E[\exp\{\alpha^s(E(U_2|A,C,Z,S=0)+\Delta_0)\}|A,C,Z,S=0]\\
&=&E(U_2|A,C,Z,S=0)+\frac{\partial}{\partial \alpha^s}\log E[\exp(\alpha^s\Delta_0)|A,C,Z,S=0]\\
&=&E(U_2|A,C,Z,S=0)+\beta^{s}(C),
\end{eqnarray*}

%\vspace{-8mm}
\noindent where $\beta^{s}(C)=\partial\log E[\exp(\alpha^s\Delta_0)|A,C,Z,S=0]/\partial \alpha^s=\partial\log E[\exp(\alpha^s\Delta_0)|C,S=0]/\partial \alpha^s$ and the last equation holds due to assumption \eqref{eq: reg_U}. Thus, $E(Y|S=s,A,C,Z)=\sum_{\tilde s\neq s}\beta^{s\tilde s}(C)\Pr(S=\tilde s|A,C,Z)+b^{s}(A,C)+\bar\tau^{s}(C)$, where $\beta^{s\tilde s}(C)=\beta^{s}(C)-\beta^{\tilde s}(C)$ and $\bar\tau^{s}(C)=E(U_2|C)+\tau^s(C)$.

%%\begin{}
%\subsection*{Proof of Proposition \ref{prop: reg}}
%%\end{appendix}
%
%
%Under model \eqref{eq: reg_Y}, we have 
%
%%\vspace{-16mm}
%\begin{eqnarray*}
%&&E(Y|S=s,A,C,Z)\\
%&=&\beta_0^s+\beta^s_aA+E(U_2|S=s,A,C,Z)+b^s(C)\\
%&=&\beta_0^s+\beta^s_aA+E(U_2|A,C,Z)\\
%&&\quad\ +\sum_{\tilde s\neq s}\{E(U_2|S=s,A,C,Z)-E(U_2|S=\tilde s,A,C,Z)\}\Pr(S=\tilde s|A,C,Z)+b^s(C)\\
%&=&\beta_0^s+\beta^s_aA+E(U_2|C)+\sum_{\tilde s\neq s}\{E(U_2|S=s,A,C,Z)-E(U_2|S=\tilde s,A,C,Z)\}\Pr(S=\tilde s|A,C,Z)\\&&+b^s(C),
%\end{eqnarray*}
%
%%\vspace{-8mm}

%\subsection*{Proof of Corollary \ref{coro: identification_causal_eff}}

Now, we show the identification of the causal effect. Under Model \eqref{eq: reg_Y_general}, we have 

$$E(Y|A=1,C,S,Z,U_2)-E(Y|A=0,C,S,Z,U_2)=b^s(1,C)-b^s(0,C).$$ We also have 

\begin{eqnarray*}
&&E(Y|A=1,C,S,Z,U_2)-E(Y|A=0,C,S,Z,U_2)\\
&=&E(Y|A=1,C,S,U_2)-E(Y|A=0,C,S,U_2)\\
&=&E(Y(1)|A=1,C,S,U_2)-E(Y(0)|A=0,C,S,U_2)\\
&=&E(Y(1)|C,S,U_2)-E(Y(0)|C,S,U_2),
\end{eqnarray*}

\noindent where the first equation is due to Assumption \ref{assumpt_exclusion}, the second and third are due to causal consistency assumption  and Assumption \ref{assumpt_ignore}, respectively. Thus, $E(Y(1)|C,S,U_2)-E(Y(0)|C,S,U_2)=b^s(1,C)-b^s(0,C)$. Integrating over $U_2$ yields that $E(Y(1)|C,S)-E(Y(0)|C,S)=b^s(1,C)-b^s(0,C)$. By Assumption \ref{assumpt_1}, the nameship mechanism $\Pr(S|A,C,Z)$ is a function of the negative control exposure $Z$. Additionally, the negative control exposure variable $Z$ only appears in the nameship mechanism model, thus, the term that involves $Z$, $\sum_{\tilde s\neq s}\beta^{s\tilde s}(C)\Pr(S=\tilde s|A,C,Z)$; and the term that does not involve $Z$, $b^{*s}(A,C)=b^s(A,C)+\bar\tau^{s}(C)$ can be identified. Hence, the causal effect can be identified.

\subsection*{Derivation of $\beta^{s\tilde s}(C)$ under Example \ref{eg: gaussian}}

Under the assumptions of Example \ref{eg: gaussian}, $\Delta_s\sim N( 0, \sigma^2)$.  Using the moment generating function, $\beta^s(C)=\partial \log \exp(\sigma^2\alpha_s^2/2)/\partial \alpha^s=\sigma^2\alpha_s$. Thus, $\beta^{s\tilde s}(C)$ can be derived.
\subsection*{Proof of Proposition \ref{prop: reg_bin}}

As shown in the proof of Proposition \ref{prop: reg_general}, assuming \eqref{eq: reg_S}--\eqref{eq: reg_U}, we have 

%\vspace{-12mm}
\begin{eqnarray*}
&&E(U_2|A,C,Z,S=s)\\
&=&E(U_2|A,C,Z)+\sum_{\tilde s\neq s}\{E(U_2|A,C,Z,S=s)-E(U_2|A,C,Z,S=\tilde s)\}\Pr(S=\tilde s|A,C,Z)\\
%&=&E(U_2|C)+\sum_{\tilde s\neq s}\{E(U|A,C,S=s)-E(U|A,C,S=\tilde s)\}\Pr(S=\tilde s|A,C)\\
&=&E(U_2|C)+\sum_{\tilde s\neq s}\{E(U_2|C,S=s)-E(U_2|C,S=\tilde s)\}\Pr(S=\tilde s|A,C,Z)\\
&=&E(U_2|C)+\sum_{\tilde s\neq s}\beta^{s\tilde s}(C)\Pr(S=\tilde s|A,C,Z).
\end{eqnarray*}

%%\vspace{-5mm}
%\noindent Thus, 
%
%%\vspace{-12mm}
%\begin{eqnarray*}
%E(U_2|A,C,S=s)&=&E(U_2|A,C,S=s)-E(U_2|A,C,S=0)+E(U_2|A,C,S=0)\\
%%&=&E(U_2|A,C)+\sum_{\tilde s\neq s}\{E(U|A,C,S=s)-E(U_2|A,C,S=\tilde s)\}\Pr(S=\tilde s|A,C)\\
%%&=&E(U_2|C)+\sum_{\tilde s\neq s}\{E(U_2|A,C,S=s)-E(U_2|A,C,S=\tilde s)\}\Pr(S=\tilde s|A,C)\\
%%&=&E(U_2|C)+\sum_{\tilde s\neq s}\{E(U_2|A,C,S=s)-E(U_2|A,C,S=\tilde s)\}\Pr(S=\tilde s|A,C)\\
%&=&\beta^{s}(C)+E(U_2|C)+\sum_{\tilde s\neq s}\beta^{s\tilde s}(C)\Pr(S=\tilde s|A,C).
%\end{eqnarray*}
%

\noindent Under Model \eqref{eq: reg_Y_bin}, we have 

%\vspace{-16mm}
\begin{eqnarray*}
&&E(Y|S=s,A,C,Z)\\
&=&E\{E(Y|U_2,S=s,A,C,Z)|S=s,A,C,Z\}\\
&=&E[\exp\{U_2+b^s(A,C)+\tau^s(C)\}|S=s,A,C,Z]\\
&=&\exp\{E(U_2|S=s,A,C,Z)+b^s(A,C)+\tau^s(C)\}E\{\exp(\Delta_s)|S=s,A,C,Z\}\\
&=&\exp\{E(U_2|C)+\sum_{\tilde s\neq s}\beta^{s\tilde s}(C)\Pr(S=\tilde s|A,C,Z)+b^s(A,C)+\tau^s(C)\}E\{\exp(\Delta_s)|S=s,C\}\\
%&=&\beta_0^s+\beta_a^sA+E(U_2|A,C)+\sum_{\tilde s\neq s}\{E(U_2|A,C,S=s)-E(U_2|A,C,S=\tilde s)\}\Pr(S=\tilde s|A,C)+b^s(C)\\
%&=&\beta_0^s+\beta_a^sA+E(U_2|C)+\sum_{\tilde s\neq s}\{E(U_2|A,C,S=s)-E(U_2|A,C,S=\tilde s)\}\Pr(S=\tilde s|A,C)+b^s(C)\\
&=&\exp\{\sum_{\tilde s\neq s}\beta^{ s\tilde s}(C)\Pr(S=\tilde s|A,C,Z)+b^s(A,C)+\bar\tau^s(C)\},
\end{eqnarray*}

%\vspace{-3mm}
\noindent where $\bar\tau^s (C)=\tau^s(C)+\log E\{\exp(\Delta_s)|S=s,C\}+E(U_2|C)$. To identify the causal contagion effect on the risk ratio scale, we have $\log\{\Pr(Y(1)=1|S,A=1,C,Z,U_2)/\Pr(Y(0)=1|S,A=1,C,Z,U_2)\}=b^s(1,C)-b^s(0,C)$, which is equivalent to $\{\Pr(Y(1)=1|S,A=1,C,Z,U_2)/\Pr(Y(0)=1|S,A=0,C,Z,U_2)\}=\exp\{b^s(1,C)-b^s(0,C)\}$, hence under Assumption \ref{assumpt_ignore} we have $\{\Pr(Y(1)=1|S,C)/\Pr(Y(0)=1|S,C)\}=\exp\{b^s(1,C)-b^s(0,C)\}$, which is the causal contagion effect. Under Assumption \ref{assumpt_1}, the nameship mechanism model $\Pr(S|A,C,Z)$ depend on $Z$, hence, following a similar argument as in the proof of Proposition \ref{prop: reg_general}, $b^s(A,C)$ and the contagion effect on the risk ratio scale can be identified.

%Assumptions \ref{assumpt_2} and \ref{assumpt_ignore}

%\subsection*{Continuous outcome with $Z$}
%\textbf{Proof of Proposition \ref{prop: reg_w_Z}:}

\clearpage

\newpage
% latex table generated in R 3.6.1 by xtable 1.8-4 package
% Sat Sep 14 15:18:24 2019
\begin{table}[ht]
\centering
\caption{Estimates, standard error and p-values of coefficients in a naive analysis without distinction among  relationships}\label{tb: naive_all_relation}\begin{tabular}{lrrrr}
  \hline
 & Est & SE &  $p$ \\ 
  \hline
  ego's BMI$_{b}$ & 0.94 & 0.01 & $<$0.01 \\ 
  alter's obese$_{b}$ & 0.24 & 0.10 & 0.01 \\ 
  ego's sex & 0.30 & 0.07  & $<$0.01 \\ 
  ego's age & -0.18 & 0.06 & $<$0.01 \\ 
  alter's sex & -0.03 & 0.07  & 0.64 \\ 
  alter's age& 0.17 & 0.06 & $<$0.01 \\ 
  age:alter's age & 0.02 & 0.02  & 0.30 \\ 
   \hline\\
\end{tabular}
\end{table}
\clearpage
\newpage
% latex table generated in R 3.3.2 by xtable 1.8-2 package
% Sat Mar 11 21:19:15 2017
\begin{table}[ht]
\centering
\caption{Estimates, standard error and p-values of coefficients in a naive analysis across different nameships: active naming ($S=1$), passive naming ($S=2$) and mutual naming ($S=3$)}\label{tb: friendNR_naive}
\begin{tabular}{lrrrrrrrrrrr}
  \hline
&\multicolumn{3}{c}{$S=1$}&&\multicolumn{3}{c}{$S=2$}&&\multicolumn{3}{c}{$S=3$}\\
 & Est & SE & $p$  &  & Est & SE & $p$  & & Est& SE & $p$ \\ 
\cmidrule{2-4}\cmidrule{6-8}\cmidrule{10-12}
%  \hline
%(Intercept) & 2.39 & 0.49 & 0.00 &  & 2.44 & 0.48 & 0.00 &  & 2.06 & 0.30 & 0.00 \\ 
 ego's BMI$_{b}$& 0.98 & 0.02 & $<$0.01 &  & 0.94 & 0.02 & $<$0.01 &  & 0.92 & 0.01 & $<$0.01 \\ 
 alter's obesity$_{b}$ & -0.16 & 0.22 & 0.47 &  & 0.37 & 0.24 & 0.11 &  & 0.33 & 0.13 & 0.01 \\ 
  sex & 0.39 & 0.17 & 0.02 &  & 0.39 & 0.17 & 0.03 &  & 0.25 & 0.08 & $<$0.01 \\ 
  age & 0.00 & 0.12 & 0.97 &  & -0.25 & 0.15 & 0.09 &  & -0.17 & 0.09 & 0.07 \\ 
 alter's sex & -0.12 & 0.16 & 0.48 &  & -0.10 & 0.17 & 0.54 &  & 0.00 & 0.08 & 0.96 \\ 
 alter's age & -0.04 & 0.14 & 0.78 &  & 0.13 & 0.14 & 0.35 &  & 0.27 & 0.09 & $<$0.01 \\ 
  age:alter's age & 0.04 & 0.05 & 0.45 &  & 0.03 & 0.05 & 0.50 &  & 0.04 & 0.04 & 0.31 \\
  
     \hline\\
\end{tabular}
\end{table}

\clearpage
\newpage
%
%\begin{table}[ht]
%\centering
%\caption{Propensity score estimates adjusted for the alter's age gender and $Z$.}
%\label{tb: propen_friendNR}
%\begin{tabular}{lrrr}
%  \hline
% & Est & SE & $p$ \\ 
%\cmidrule{2-4}%Intercept & 1.47 & 0.18 & 0.00 \\ 
%  alter's obesity$_{b}$ & $-$0.01 & 0.01 & 0.41 \\
%  sex  & $-$0.16 & 0.06 & 0.01 \\ 
%  age & $-$0.45 & 0.06 & $<$0.01 \\ 
%  alter's sex & 0.04 & 0.08 & 0.63 \\
%  alter's age & $-$0.08 & 0.06 & 0.14 \\
%  age:alter's age & 0.00 & 0.00 & 0.10 \\ 
%  $Z$ & 0.01 & 0.00 & 0.02 \\ 
% \hline\\
%\end{tabular}
%\end{table}

\clearpage
\newpage

\begin{table}[ht]
\centering
\caption{Nameship mechanism estimates adjusted for alter's age gender and $Z$.}
\label{tb: propen_friendNR_new}
\begin{tabular}{lrrr}
  \hline
\multicolumn{4}{c}{Ego model}\\
   \hline
 & Est & SE & $p$ \\ 
\cmidrule{2-4}
  alter's BMI$_{b}$& 0.01 & 0.01 & 0.39 \\ 
  ego's  sex & 0.09 & 0.08 & 0.26 \\ 
  ego's  age & -0.24 & 0.07 & $<$0.01 \\ 
  alter's sex & -0.08 & 0.09 & 0.35 \\ 
  alter's age & -0.41 & 0.07 & $<$0.01 \\ 
\hline
\multicolumn{4}{c}{Alter model}\\
   \hline
 & Est & SE & $p$ \\ 
\cmidrule{2-4}
  ego's  BMI & 0.02 & 0.01 & 0.08 \\ 
  alter's sex & 0.06 & 0.09 & 0.46 \\ 
  alter's age & -0.48 & 0.07 & $<$0.01 \\ 
  ego's  sex & 0.21 & 0.09 & 0.02 \\ 
  ego's age & -0.18 & 0.07 & 0.01 \\ 
  $Z$ & -0.03 & 0.01 & $<$0.01 \\
\hline\\
   \end{tabular}
\end{table}

\clearpage
\newpage

\begin{table}[ht]
\centering
\caption{Estimates, sandwich standard error and p-values of coefficients in homophily-adjusted analysis with an negative control exposure  variable $Z$ across different nameships: active naming ($S=1$), passive naming ($S=2$) and mutual naming ($S=3$)}\label{tb: friendNR_2stage_new_sandwich}
\begin{tabular}{lrrrrrrrrrrr}
  \hline
&\multicolumn{3}{c}{$S=1$}&&\multicolumn{3}{c}{$S=2$}&&\multicolumn{3}{c}{$S=3$}\\
 & Est & SE &$p$ &  & Est & SE &$p$ & & Est & SE & $p$ \\ 
\cmidrule{2-4}\cmidrule{6-8}\cmidrule{10-12}
%  \hline
%(Intercept) & 59.78 & 10.97 & 0.00 &  & -198.20 & 12.84 & 0.00 &  & 127.84 & 3.32 & 0.00 \\ 
 ego's BMI$_{b}$ & 1.00 & 0.03 & $<$0.01 &  & 0.93 & 0.03 & $<$0.01 &  & 0.93 & 0.02 & $<$0.01 \\
 alter's obesity$_{b}$  & -0.24 & 0.33 & 0.47 &  & 0.52 & 0.34 & 0.13 &  & 0.21 & 0.19 & 0.27 \\
  sex & 0.82 & 0.35 & 0.02 &  & 0.30 & 0.36 & 0.39 &  & 0.22 & 0.18 & 0.23 \\
  age & -0.74 & 0.58 & 0.21 &  & -0.07 & 0.74 & 0.93 &  & 0.30 & 0.55 & 0.58 \\ 
  alter's sex & -0.25 & 0.33 & 0.45 &  & -0.17 & 0.27 & 0.52 &  & 0.11 & 0.13 & 0.37 \\ 
 alter's age & -1.42 & 0.92 & 0.12 &  & 0.37 & 1.19 & 0.76 &  & 0.80 & 0.76 & 0.29 \\
 age:alter's age & 0.08 & 0.17 & 0.63 &  & 0.14 & 0.24 & 0.57 &  & 0.23 & 0.17 & 0.18 \\ 
$\beta^{ss_1}$& -10.93 & 5.88 & 0.06 &  & -2.93 & 14.91 & 0.84 &  & -10.97 & 9.64 & 0.26 \\ 
$\beta^{ss_2}$  & -8.81 & 11.67 & 0.45 &  & 3.45 & 12.49 & 0.78 &  & -3.22 & 7.00 & 0.65 \\ 
$\beta^{ss_3}$& -20.67 & 10.57 & 0.05 &  & 1.02 & 4.66 & 0.83 &  & 3.80 & 2.98 & 0.20 \\  
    \hline\\
\end{tabular}
\end{table}
\clearpage
\newpage

\clearpage

\clearpage

\label{lastpage}

\end{document}